\documentstyle[11pt,epsfig,here,mine]{article}
\begin{document}

\begin{center}
{\Large
{\it
Submitted to the Proceedings of International Workshop\\
XXXIInd Rencontres de Moriond -- Very High Energy Phenomena\\
(Les Arcs, Savoie, France, January 18 -- 25, 1997)\\ 
}}
\end{center}

\vspace*{2.5cm}
\title{THE BAIKAL EXPERIMENT: STATUS REPORT}

\author{
V.A.BALKANOV$^2$, I.A.BELOLAPTIKOV$^7$, L.B.BEZRUKOV$^1$, B.A.BORISOVETS$^1$, 
N.M.BUDNEV$^2$, A.G.CHENSKY$^2$, I.A.DANILCHENKO$^1$, ZH.-A.M.DJILKIBAEV$^1$, 
V.I.DOBRYNIN$^2$, G.V.DOMOGATSKY$^1$, A.A.DOROSHENKO$^1$, S.V.FIALKOVSKY$^4$,
O.N.GAPONENKO$^2$, A.A.GARUS$^1$, S.B.IGNAT$'$EV$^3$, A.KARLE$^8$, 
A.M.KLABUKOV$^1$, A.I.KLIMOV$^6$, S.I.KLIMUSHIN$^1$, A.P.KOSHECHKIN$^1$, 
V.F.KULEPOV$^4$, L.A.KUZMICHEV$^3$, B.K.LUBSANDORZHIEV$^1$, T.MIKOLAJSKI$^8$,
M.B.MILENIN$^4$, R.R.MIRGAZOV$^2$, A.V.MOROZ$^2$, N.I.MOSEIKO$^3$, 
S.A.NIKIFOROV$^2$, E.A.OSIPOVA$^3$, A.I.PANFILOV$^1$, YU.V.PARFENOV$^2$, 
A.A.PAVLOV$^2$, D.P.PETUKHOV$^1$, P.G.POKHIL$^1$, P.A.POKOLEV$^2$, 
M.I.ROZANOV$^5$, V.YU.RUBZOV$^2$, I.A.SOKALSKI$^1$, CH.SPIERING$^8$, 
O.STREICHER$^8$, B.A.TARASHANSKY$^2$, T.THON$^8$, D.B.VOLKOV$^2$, 
CH.WIEBUSCH$^8$, R.WISCHNEWSKI$^8$ 
}

\address{
1 - Institute  for  Nuclear  Research,  Russian  Academy  of   Sciences
(Moscow); \mbox{2 - Irkutsk} State University (Irkutsk); \mbox{3 - Moscow}
State University (Moscow); \mbox{4 - Nizhni}  Novgorod  State  Technical
University  (Nizhni   Novgorod); \mbox{5 - St.Petersburg} State  Marine
Technical  University  (St.Petersburg); \mbox{6 - Kurchatov} Institute
(Moscow); \mbox{7 - Joint} Institute for Nuclear Research (Dubna);
\mbox{8 - DESY} Institute for High Energy Physics (Zeuthen) 
}

\author{
{\large presented by I.SOKALSKI}
}

\address{
          E-mail: sokalski@pcbai10.lpi.msk.su
}

\maketitle\abstracts{
\looseness=-1
We review the present status of the Baikal Neutrino Project. The construction
and performance of the large deep underwater Cherenkov detector for muons and 
neutrinos, {\it NT-200}, which is currently under construction in Lake Baikal 
are described. Some results obtained with the first stages of {\it NT-200} --
{\it NT-36} (1993-95), {\it NT-72} (1995-96) and {\it NT-96} (1996-97) -- are 
presented, including the first clear neutrino candidates selected with 1994 
and 1996 data.
}

\section{Introduction}

\looseness=-2
The possibility to build a neutrino telescope in Lake Baikal was investigated 
since 1980, with the basic idea to use -- instead of a ship -- the winter ice 
cover as a platform for assembly and deployment of instruments~\cite{chud}. 
After first small size tests, in 1984-90 single-{\it string} arrays equipped 
with 12 -- 36 PMTs ({\it FEU-49} with flat 15 cm photocathode) were deployed 
and operated via a shore cable~\cite{girl}. The total life time for these 
"first generation detectors" made up 270 days. On the methodical side, 
underwater and ice technologies were developed, optical properties of the 
Baikal water as well as the long-term variations of the  water luminescence 
were investigated in great details. For the Baikal telescope site the 
absorbtion length for wavelength between 470 and \mbox{500 nm} is about 20 m, 
typical value for scattering length is 15 m~\footnote{
\looseness=-2
               Sometimes the {\it effective scattering length}
               $L_{eff} = L_{scatt} / (1 - \langle \cos \theta \rangle)$ is
               used to characterize the relative merits of different sites for
               neutrino telescopes~\cite{Price}. With $L_{scatt}$= 15\,m  and 
               $\langle \cos \theta \rangle$ = 0.95  one
               obtains \mbox{$L_{eff}$ = 300\,m} for the Baikal site.
              }
with mean cosine of the scattering angle being close to 0.95 (see~\cite{APP} 
and refs. therein). Since 1987, a "second generation detector" with the 
capability to identify muons from neutrino interactions was envisaged. 
Tailored to the needs of the Baikal experiment, a large area hybrid phototube 
{\it QUASAR}~\cite{Pylos} with hemispherical photocatode of 37 cm diameter and
a time resolution of better than 3\,nsec was developed to replace {\it FEU-49}.
According to the approximate number of PMTs this detector was named 
{\it NT-200} -- Neutrino Telescope with 200 PMTs~\cite{baikal}.  With estimated
effective area of about 2300\,m$^2$ and 8500\,m$^2$ for 1-TeV and 100-Tev 
muons, respectively, it is a first stage of a future full-scale telescope, 
which will be built stepwise, via intermediate detectors of rising size and 
complexity. 

\section{The Baikal Neutrino Telescope {\it NT-200}}

\vspace{-1mm}
\looseness=-1
The Baikal Neutrino Telescope (Fig.1) is being deployed in Lake Baikal, 
\mbox{3.6 km} from shore at a depth of \mbox{1.1 km}. It will consist of 192 
optical modules (OMs). The umbrella-like frame carries the 8 strings with the 
detector components. The OMs are grouped in pairs along the strings. The 
pulses from two PMTs of a pair after \mbox{0.3 {\it p.e.}} discrimination are 
fed to a coincidence with \mbox{15 ns} time window. A pair defines a 
{\it channel}. A {\it muon-trigger} is formed by the requirement of 
\mbox{$\geq N$ {\it hits}} (with {\it hit} refering to a channel) within 
\mbox{500 ns}. $N$ is typically set to the value \mbox{3 or 4.} Then digitized
amplitudes and times of all hit channels are sent to the shore. A separate 
{\em monopole trigger} system searches for time patterns characteristic for 
slowly moving objects.

\begin{figure}[H]
\centering
\mbox{\epsfig{file=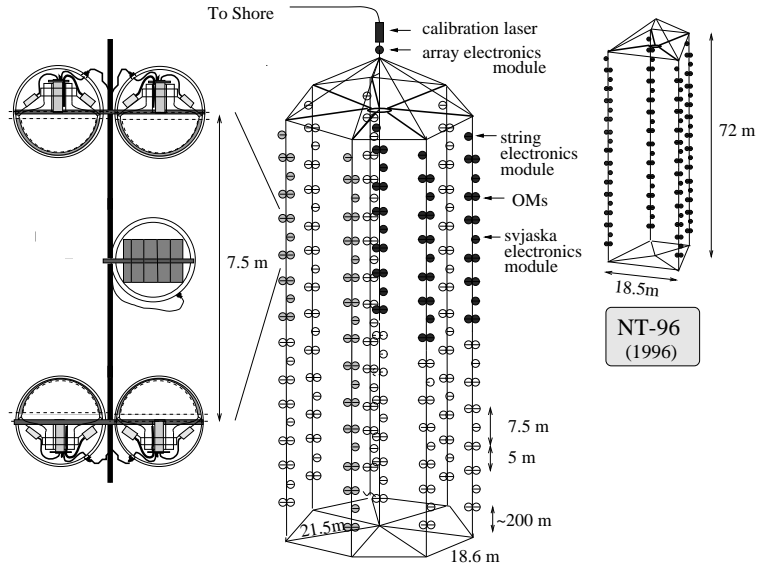,height=10cm,angle=-90}}
%%%\hspace{0.0cm}
%%%\parbox[b]{4.5cm}{\caption [2] {
\parbox{15cm}{\caption [1] {
\frenchspacing
Schematic view of {\em NT-200}. Components deployed in 1993 are in black, 
those added in 1995 in grey. Top right: the strongly modified array deployed 
in 1996.
\nonfrenchspacing
}}
\end{figure}

\looseness=-1
In April 1993, the first part of {\it NT-200}, the detector {\it NT-36} with 
36 OMs at 3 short strings, was put into operation and took data up to March 
1995. A 72-OMs array, {\it \mbox{NT-72}}, run in \mbox{1995-96}. In 1996 it 
was replaced by the four-string array {\it NT-96}. Summed over 700 days 
effective life time, $3.2\cdot 10^{8}$ muon events have been collected with 
\mbox{{\it NT-36, -72, -96}}. Since \mbox{April 6,} 1997, {\it NT-144}, a 
six-string array with 144 OMs, is taking data in Lake Baikal.

\section{Track Reconstruction}

\vspace{-1mm}
In contrast with a typical underground detector, it is impossible to determine
co-ordinates for some clearly visible points which would belong to a track of 
a particle crossing an underwater array because it represents a lattice of OMs
with large distances between them. The parameters of a single muon track have 
to be determined~\cite{Reco} by minimizing

\vspace{-4mm}
\begin{displaymath}
S^2_t = \sum_{i=1}^{N_{hit}} (T_i(\theta, \phi, u_0, v_0, t_0)
    - t_i)^2 / \sigma_{ti}^2.
\end{displaymath}
\vspace{-4mm}

\noindent
\looseness=-1
Here, $t_i$ are the measured times and $T_i$ the times expected for a given 
set of track parameters. $N_{hit}$ is the number of hit channels, 
$\sigma_{ti}$ are the timing errors. A set of parameters defining a straight 
track is given by $\theta$ and $\phi$ -- zenith and azimuth angles of the 
track, respectively, $u_0$ and $v_0$ -- the two coordinates of the track point 
closest to the center of the detector, and $t_0$ -- the time the muon passes 
this point. For the results given here we do not include an amplitude term 
$S_a^2$ analog to $S_t^2$ in the analysis, but use the amplitude information 
only to calculate the timing errors  $\sigma_{ti}$ in the denominator of the 
formula above. Only events fulfilling the condition  \mbox{``$\geq 6$} hits at
\mbox{$\geq 3$} strings`` are selected for the track reconstruction procedure 
which consists of the following steps:

\noindent
{\bf 1.} A preliminary analysis includes several causality criteria rejecting 
events which violate the model of a naked muon. After that, a 0-th 
approximation of $\theta$ and $\phi$ is performed.
\vspace{1mm}

\noindent
\looseness=-2
{\bf 2.} The $\chi^2$ minimum search, based on the model of a naked muon and 
using only time data.
\vspace{1mm}

\noindent
{\bf 3.} Quality criteria to reject most badly reconstructed events.
\vspace{1mm}

\looseness=-1
We have developed a large set of pre-criteria as well as quality criteria to 
reject misreconstructed events. Most of these criteria are not independent of 
each other.  Furthermore, the optimum set of criteria turned out to depend on 
the detector configuration. The causality criteria refer to time differences  
between channels. {\it E.g.}, one requests that each combination of two 
channels $i,j$ obeys the condition $c~|dt_{ij}| < n~|dx_{ij}| + c~\delta t$,
where $dt_{ij}$ and $dx_{ij}$ are time differences and distances between 
channels $i$ and $j$, respectively and $n = 1.33$ is refraction coefficient 
for the water. The term $\delta t =$ 5\,nsec accounts for time jitter. Some of
the most effective quality criteria are, {\it e.g.}, upper limits on 
parameters like the minimum $\chi^2$, the probability $P_{nohit}$ of non-fired 
channels not to respond to a naked muon and probability $P_{hit}$ of fired 
channels to respond to a naked muon. 

\section{Selected Results}

\subsection{Atmospheric Muons Vertical Flux}

Muon angular distributions are well described by MC expectations. Converting 
the measured angular dependence obtained with standard reconstruction applied
to {\it NT-36} data~\cite{APP} into a depth dependence of the vertical flux, 
good agreement with theoretical predictions is observed (Fig.2).

\subsection{Upper Limits on the Flux of Monopoles Catalyzing Baryon Decay}

For certain regions of the parameter space in $\beta$ (monopole velocity) and 
$\sigma _{c}$ (catalysis cross section), GUT monopoles would cause sequential 
hits in individual channels in time windows of $10^{-4}$--$10^{-3}$ sec. 
Having searched for such enhanced counting rates in the 1993 data obtained 
with {\it NT-36} array we deduce upper limits for the flux of monopoles 
catalyzing the decay of protons with cross section~\cite{ruba} 
\mbox{$\sigma _{c}=0.17\cdot \sigma _{0}\cdot \beta ^{-2}$}. Our limits 
\mbox{(90 $\%$ CL)} are shown in Fig.3  together with our earlier 
results~\cite{girl}, limits from IMB~\cite{imbmon} and 
Kamiokande~\cite{kammon} and with the astrophysical Chudakov-Parker 
limit~\cite{parker}. A limit of \mbox{$4\cdot 10^{-16}$ cm$^{-2}$ s$^{-1}$} 
has been obtained by the Baksan Telescope~\cite{bakmon} for 
\mbox{$\beta > 2 \cdot 10^{-4}$}. Further progress is expected with the next 
stages of {\it NT-200}.

\vspace{-7mm}
\begin{figure}[H]
\mbox{\epsfig{file=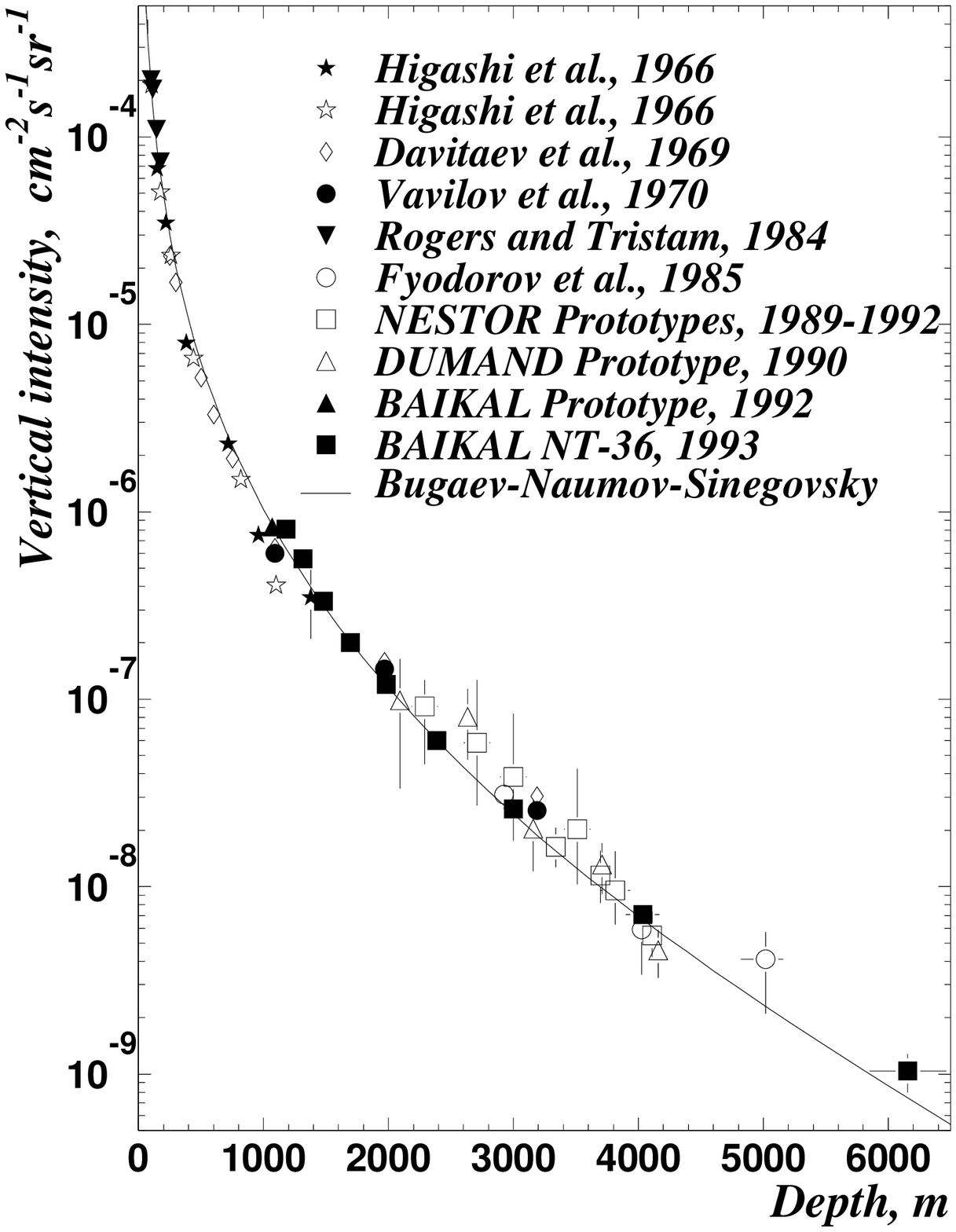,width=7.5cm,height=8.7cm}}
\mbox{\epsfig{file=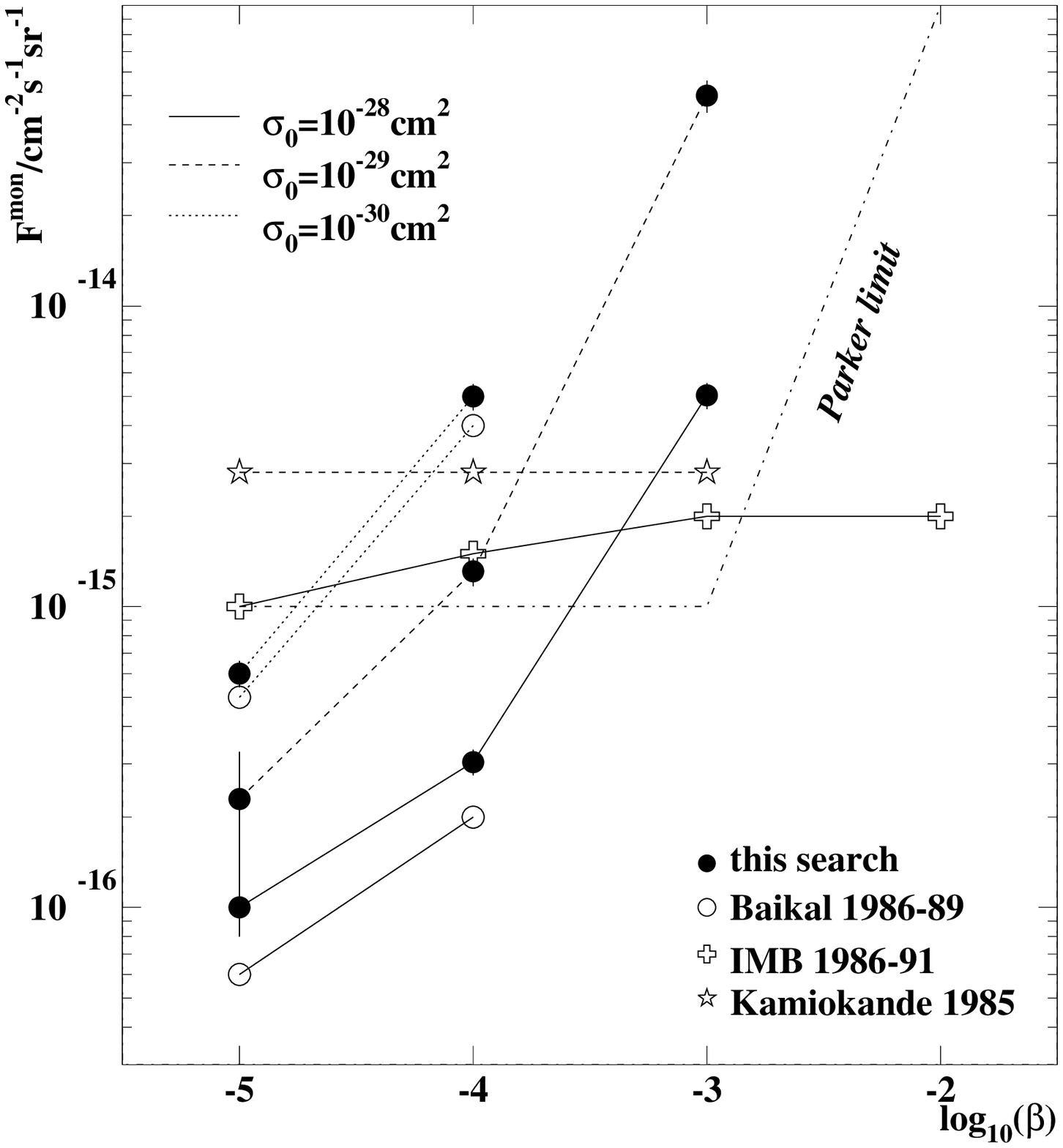,width=7.5cm,height=8.0cm}}
     \parbox[t]{7.3cm}{
          \caption [2] {
  Vertical muon flux, $I_{\mu}(cos\theta=1)$, vs. water depth $L$. The nine
  {\it NT-36} values (full triangles) are calculated for $\cos\theta$\,=\, 
  0.2 to 1.0 in steps of 0.1. The curve represents theoretical 
  predictions~\cite{bug}. The other data points are taken from 
  refs.~\cite{vertical}. 
}}
\hspace{.4cm}
     \parbox[t]{7.3cm}{\caption [3]{
   \mbox{Upper limits (90 $\%$ CL)} on the natural flux of magnetic monopoles 
   catalyzing barion decqay versus their velocity $\beta$, for different 
   parameters $\sigma_o$, see text.
}}
\end{figure}

\subsection{Separation of Neutrino Events with Standard Track Reconstruction}

\vspace{-1mm}
\looseness=-1
The most obvious way to select events from the lower hemisphere (which 
dominantly are due to atmospheric neutrino interactions in the ground or water
below the array) is to perform the full spatial track reconstruction (see 
Sec. 3) and select the events with negative $\theta$ values. Taking into 
account that the flux of downward muons is about 6 orders of magnitude larger 
than the flux of upward muons, the reconstruction procedure should be 
performed extremely thoroughly. Even if very small fraction of downward 
atmospheric muons is misreconstructed as up-going ones, it forms an essential 
background. Due to small value of $S/N$ ratio (where $S$ is counting rate of 
upward neutrino induced events and $N$ is counting rate of downward 
atmospheric muons which are reconstructed as upward events), it is impossible 
to observe clear neutrino signal with {\it NT-36} and {\it NT-72} data and the
current level of standard reconstruction procedure. MC calculations indicate 
the essetially better characteristics for {\it NT-96} detector which can be 
considered as a neutrino telescope in the wide region of $\theta$. The 
analysis of the {\it NT-96} data aimed to search for the upward neutrino 
induced muons using the standard reconstruction procedure is presently in 
progress. The reconstruction procedure is tuned in this first analysis to a 60
degree half-aperture cone around the opposite zenith. 1.2 neutrino induced 
events per week are expected from MC calculations. We have analyzed 12.9 days 
data sample and selected 3 neutrino candidates with an expected number of 2.3.
One of them is presented in Fig.4. We hope that further analysis of 
{\it NT-96} and {\it NT-144} data will confirm our capability to select the 
neutrino induced events over the background of fake events from downward 
atmospheric muons.  

\begin{figure}[H]
%%%\hspace{-1.6cm}
\centering
\mbox{\epsfig{file=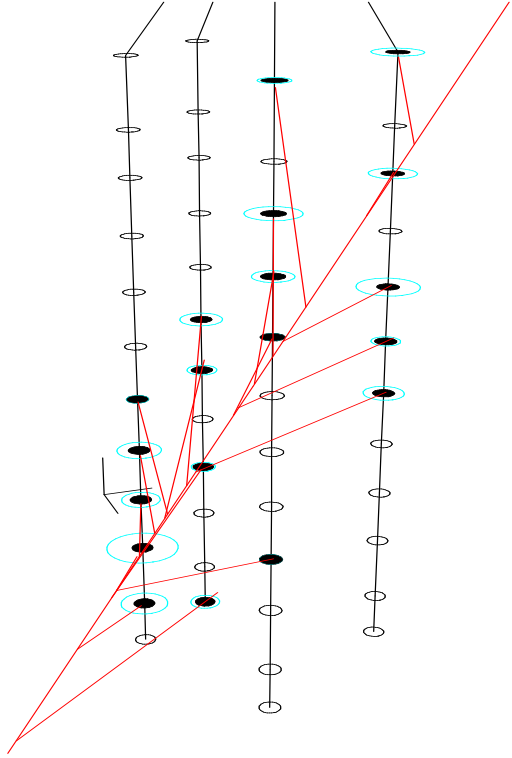,height=10cm,width=12cm}}
%%%\hspace{-2.4cm}
%%%\parbox[b]{8.1cm}{\caption [4]
\parbox{15cm}{\caption [4]
{\frenchspacing
An {\it NT-96} event reconstructed as an upward muon. The thick line gives the
reconstructed path, thin lines pointing to the channels mark the path of 
Cherenkov light as given by the fit. The areas of the circles are proportional
to the measured amplitudes. Some characteristics of given event (see text for 
explanation): Array: {\it NT-96};
date: 19.04.96;
number of hit channels 19;
$P_{hit}$:  0.38;
$P_{nohit}$: 0.75;
$\chi^2_{time}/NDF$: 0.48;
$\chi^2_{ampl}/NDF$: 1.2;
maximal distance between hit channels: 72.6 m; 
reconstructed $\theta$: -152.7 degree;
reconstructed $\phi$: 253.5 degree.
}}
\end{figure}

\subsection{Search for Nearly Upward Moving Neutrinos}

\looseness=-1
To identify nearly vertically upward muons with energies below 1 TeV (as 
expected, {\it e.g.}, for muons generated by neutrinos resulting from dark 
matter annihilation in the core of the Earth), full reconstruction is found to
be not neccessary~\cite{ourneu}. Instead, separation criteria can be applied 
which make use of two facts: firstly, that the muons searched for have the 
same  vertical direction like the string; secondly, that low-energy muons 
generate mainly direct Cherenkov light and, consequently, are not visible over
large distances and should produce a clear time and amplitude pattern in the 
detector. We have choosen the following separation criteria: 

\noindent
{\bf 1.} 3 down-facing and at least 1 up-facing channels at one string
exclusively must be hit with time differences between any 2 hit channels $i$
and $j$ obeying the condition 

\begin{center}
\mbox{$|(t_{i}-t_{j})-(T_{i}-T_{j})| < 20 ns$},
\end{center}

\noindent
where
$t_{i}(t_{j})$ are the measured times, $T_{i}(T_{j})$ are the times expected
for minimal ionizing, up-going vertical muons.
\vspace{1mm}

\noindent
{\bf 2.} The signals of down-facing channels should be significantly larger
than those of upward facing channels. Any combination of oppositely directed
hit channels must obey the inequality

\begin{center}
\mbox{$(A_{i}(\downarrow)-A_{j}(\uparrow))/(A_{i}(\downarrow)+A_{j}(\uparrow))>
0.3$} 
\end{center}

\noindent
with $A_{i}(\downarrow)$ and $A_{j}(\uparrow))$ being the amplitudes of
channels $i$ and $j$ facing down and up, respectively.
\vspace{1mm}

\noindent
{\bf 3.} All signals of down facing channels must exceed a minimum value of
\mbox {4 {\it p.e.}} (this prevents that the previous cut is dominated by low
amplitude events and, consequently, by fluctuations of the few-photoelectron
statistics).
\vspace{1mm}

\looseness=-1
The analysis presented here is based on the data taken with {\it NT-36} during
the period \mbox{April 8,} 1994, to March 1, 1995 (212 days of detector life 
time). Upward-going muon candidates were selected from a total of 
\mbox{$8.33 \cdot 10^{7}$} events recorded during this period by the 
muon-trigger \mbox{$\geq 3$.} We found 2 candidates (Fig.5) passing our cuts
with an expected number of 1.2 events from atmospheric neutrinos as obtained 
from MC (the samples fulfilling trigger conditions {\it 1, 1-2} and {\it 1-3} 
contain 131, 17 and 2 events, respectively). Our preliminary estimates for 
these candidates indicate the probability to be fake events to be equal to 
\mbox{$10^{-2} - 10^{-1}$.} Therefore we consider them as clear neutrino 
candidates. 

\begin{figure}[H]
\centering
\mbox{\epsfig{file=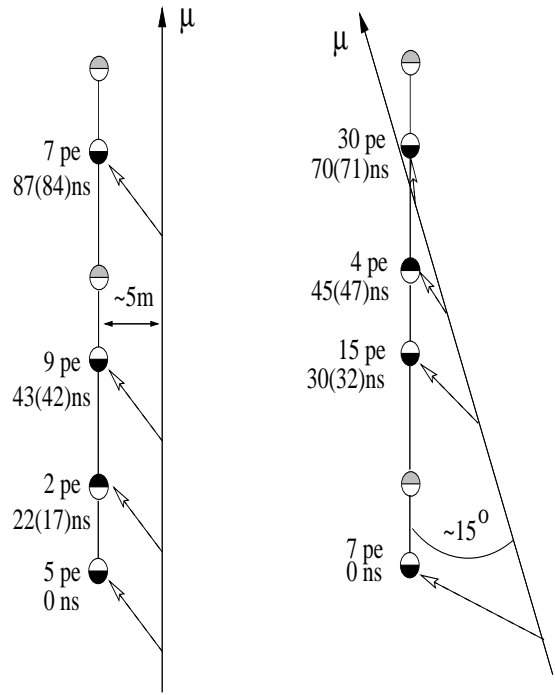,height=7.2cm,width=9.2cm,angle=-90}}
%%%\hspace{0.0cm}
%%%\parbox[b]{6.5cm}{\caption [5] {
\parbox{15cm}{\caption [5] {
\frenchspacing
The two neutrino candidates. The hit PMT pairs (channels) are marked by black.
Numbers give the measured amplitudes (in photoelectrons) and times with 
respect to the first hit channel. Times in brackets are those expected for a 
vertical going upward muon (left) and an upward muon passing the string under
\mbox{$15^o$} (right).
\nonfrenchspacing
}}
\end{figure}

\noindent
Regarding them as atmospheric neutrino events, an upper limit of 
$1.3 \cdot 10^{-13}$ muons/cm$^{2}$/sec \mbox{(90 \% CL)} in a cone with 15 
degree half-aperture around the opposite zenith is obtained for muons 
generated by neutrinos due to neutralino annihilation in the core of the 
Earth. The limit corresponds to muon energies greater than 
\mbox{$\approx 6$ GeV}. This is still an order of magnitude higher than the 
limits obtained by Kamiokande~\cite{kamneu}, Baksan~\cite{bakneu} and 
MACRO~\cite{macneu} but considerable progress is expected from the further
analysis of {\it NT-72, -96} and {\it -144} data. The {\it NT-36} effective 
area of muons fulfilling our cuts is \mbox{$S=50$ m$^{2}$/string}. A rough 
estimate of the effective area of the full-scale \mbox{{\it NT-200}} with 
respect to vertically upward going muons gives 
\mbox{$S \approx 400-800$ m$^{2}$.} At the moment the data obtained with 
{\it NT-96} array analyzed in a similar way should yield approximately 10 
neutrino candidates. This analysis is expected to confirm the method.  

\vspace{-1mm}
\section{The Next Steps}

\vspace{-1mm}
On April 6, 1997, a six-string array with 144 optical modules, {\it NT-144},
was put into operation. By April 30, 1997 it has collected 
$ \approx 2 \cdot 10^{7}$ muons. We plan to complete the {\it NT-200} array in
April, 1998.

\end{document}